\documentclass{ifacconf}

\usepackage{multirow}
\usepackage{amsfonts}
\usepackage{url}

\usepackage{stfloats}

\usepackage{amsmath}
\usepackage{amssymb}
\usepackage{enumerate}
\usepackage{definice}
\usepackage{color}
\usepackage{booktabs,hhline} 
\usepackage{cool}
\usepackage{natbib}
\usepackage{mathtools}
\usepackage{makecell}
\usepackage{xspace}
\usepackage{tikz}
\usetikzlibrary{arrows,arrows.meta,backgrounds,calc,decorations.markings,decorations.pathreplacing,fadings,fit,intersections,positioning,shadows,shapes.misc,shapes}
\usepackage{pgfplots,pgfplotstable}
\pgfplotsset{compat=newest}

\newcommand{\MATLAB}{\textsc{Matlab}\xspace}

\usepackage[scr=boondoxo,scrscaled=1.05]{mathalfa}

\def\bfD{\mathbf D}

\def\bfG{\mathbf G}

\def\bfP{\mathbf P}

\def\bfS{\mathbf S}

\def\bff{\mathbf f}
\def\bfg{\mathbf g}

\def\bfu{\mathbf u}
\def\bfv{\mathbf v}
\def\bfw{\mathbf w}
\def\bfx{\mathbf x}
\def\bfy{\mathbf y}
\def\bfz{\mathbf z}

\def\bfDelta{\mathbf \Delta}
\def\bfSigma{\mathbf \Sigma}

\def\calB{\mathcal{B}}

\def\calN{\mathcal{N}}

\def\calS{\mathcal{S}}

\def\calU{\mathcal{U}}

\def\calN{\mathbfcal{N}}

\DeclareMathAlphabet\mathbfcal{OMS}{cmsy}{b}{n}

\def\real{\mathbb{R}}

\def\mean{\mathsf{E}}

\def\cov{\mathsf{cov}}
\def\abs{\mathsf{abs}}

\def\l{^{L}}

\usepackage{graphicx}      
\usepackage{natbib}        

\def\nu{c}
\def\nz
\def\l{^{L}}

\newcommand{\tightoverset}[2]{%
	\mathop{#2}\limits^{\vbox to -.5ex{\kern-0.75ex\hbox{$#1$}\vss}}}

\begin{document}
	\begin{frontmatter}
		
  \title{Data-Augmented Numerical Integration in State Prediction: Rule Selection \hspace{-0mm}}
		
		
        \author[First]{J. Dun\'{i}k}
        \author[First]{L. Král}
        \author[First]{J. Matoušek}
        \author[First]{O. Straka}
        \author[First]{M. Brandner$^*$} 
        
        \address[First]{Dept. of Cybernetics, $^{**}$Dept. of Mathematics\\ Univ. of West Bohemia, Univerzitní 8, 306 01 Pilsen, Czech Republic\\ e-mails: \{dunikj,ladkral,matoujak,straka30\}@kky.zcu.cz, brandner@kma.zcu.cz}

		\begin{abstract} 
            This paper deals with the state prediction of nonlinear stochastic dynamic systems. The emphasis is laid on a solution to the integral Chapman-Kolmogorov equation by a deterministic-integration-rule-based point-mass method. A novel concept of reliable \textit{data-augmented}, i.e., mathematics- and data-informed, integration rule is developed to enhance the point-mass state predictor, where the trained neural network (representing \textit{data} contribution) is used for the selection of the best integration rule from a set of available rules (representing \textit{mathematics} contribution). The proposed approach combining the best properties of the standard mathematics-informed and novel data-informed rules is thoroughly discussed.\footnote{\copyright 2024 the authors. This work has been accepted to IFAC for publication under a Creative Commons Licence CC-BY-NC-ND}
		\end{abstract}
		
		\begin{keyword}
			State estimation, Neural network, Numerical integration, Nonlinear predictors, Bayesian methods, Stochastic
	systems.
		\end{keyword}
		
	\end{frontmatter}

	\section{Introduction}
State estimation of nonlinear discrete-time stochastic dynamic systems from noisy or incomplete measurements has been a subject of considerable research interest for the last seven decades. Prediction can be understood as a task of estimating the future state with respect to the last available measurement. Thus, this task is essential in many applications ranging from signal processing and predictive control through navigation and economy to weather forecast. State prediction is also an inherent part of any filtering or smoothing algorithm 
\citep{Sa:13},
\citep{ChSaGo:15}.

In this paper, the stress is laid on the state prediction of a system modeled by a nonlinear discrete-in-time stochastic dynamic model. In particular, the emphasis is put on the Bayesian approach, where the predictive (conditional) probability density function (PDF) of the state is calculated. A general solution to the prediction is given by the \textit{integral} Chapman-Kolmogorov equation (CKE). 

\subsection{Related Work}
The CKE is exactly solvable for a limited set of dynamic models, for which the linearity and Gaussianity are usually common factors \citep{Sa:13,Stro:71}. In other cases, i.e., for nonlinear or non-Gaussian models, an approximate solution to the CKE is employed. The approximate solutions to the CKE for a nonlinear model either \textit{(i)} linearize the model or apply the numerical integration rules, or \textit{(ii)} utilize a machine learning method for approximation of the numerical solution or the prediction. 

\subsubsection{Mathematical-driven Methods}

The \textit{first} group of approximate methods is based on the linearization of the involved nonlinear functions in the dynamics and assumption of a Gaussian PDF of all related random variables, which allows to apply the solution to the CKE known for the linear and Gaussian model. Often, linearization techniques based on the Taylor expansion, Stirling's interpolation, or statistical linearization are used, and the resulting prediction is in the form of the mean and covariance matrix 
\citep{JuUhl:04,SiDu:09,Sa:13}.
The \textit{second} group solves the integral CKE by various numerical integration rules (IRs). In the literature, two groups of methods can be found, namely density-specific and grid-based. Whereas the density-specific methods rely on the assumption of certain forms of the integrated PDFs (e.g., Gaussian, Student's-t) \citep{JuUhl:04,SiDu:09,DuStSiBl:15}, the grid-based methods cover the state-space by a grid of deterministically or randomly placed points used for the integral solution. The popular integration methods are represented by, e.g., Monte-Carlo methods or the midpoint rule  \citep{Stro:71,DoFrGo-book:01,ClFeFe:09,Sa:13},
\citep{DuStMaBl:22}. To achieve higher accuracy, the Richardson extrapolation or the Newton-Cotes formulas, such as a trapezoid or Simpson rule, can be applied \citep{St:02}. For multiple-dimensional integration, the IRs for the specific dimensions are proposed as an alternative to a combination of one-dimensional rules \citep{Mi:60}. Besides the above-mentioned deterministic IRs with the equidistantly spaced integration nodes, the adaptive IRs with non-equidistantly spaced grid can also be used. The adaptive rules might provide better performance \citep{Pr:07}, but usually at the cost of higher computational complexity. Note that the methods of this group are also suitable for linear models with a non-Gaussian prior PDF. 

Application of these integration methods has led to a plethora of predictors ranging from rather simple ones utilizing the prediction step of the local filters, such as the extended or unscented Kalman filter, to more advanced global ones being based on the particle or point-mass filters \citep{SiDu:09,Sa:13}. The methods have been thoroughly analyzed with regard to convergence and integration errors.

\subsubsection{Data-driven Methods}
Recently, the advancement in machine learning has enabled the design of the \textit{second} group of integration methods, i.e., of the \textit{data-driven} integration rules \citep{ZhYaHu:06,LlIrAh:20,Yo:21,AlGaYa:23}. These rules solve the integral relations using the pre-trained neural networks (NN). Indeed, the integral can be seen as a transformation of an integrand to an integration result, and as any other transformation (or function), it can be completely substituted by the NN (end-to-end approach). Note that the approaches considered have a major limitation: the assumption of fixed integration. As such, they are \textit{not} suitable in the considered problem of the CKE solution since the integrand varies at each time instant of state estimation (due to the time evolution of the PDF). Besides the area of numerical integration, the NNs have been directly used in the state estimator design for more than two decades. Originally, the NNs were used for system modeling with or without respecting the physics behind \citep{GoMe:08,BaVeBaSh:20,Sc:21}. These models were then used in the estimation or prediction algorithms discussed above. Then, the NNs have been used directly at the level of estimation algorithms, i.e., to replace the prediction or estimation algorithms with the NN without the need for a model knowledge \citep{ZhYiLiJuKo:19,ShAlIqMoOt:21,HeLe:18,ZhYiLiJuKo:19,TiZhBiQiQu:20,AlGaYa:23}. Compared to the mathematics-driven methods, the NN-based approaches do not require any deep insight into the solved task. On the other hand, the NN has to be well (or life-long) trained to provide reliable results, which cannot be analytically assessed in terms of integration output properties. NN training is typically time and computationally demanding.

\subsection{Motivation and Paper Contribution}
Irrespectively to which method is used for an approximate CKE solution, the calculated predictive PDF is inherently associated with an \textit{integration error}, and the selection of the suitable IR with required properties remains \textit{difficult} and \textit{problem specific} task with no globally valid guidelines.  Ultimately, the IR selection is left to the predictor designer or user, who is expected to possess a deeper insight into the solved task and knowledge of a wide range of data and mathematics-driven numerical methods.

The \textit{goal} of the paper is to propose, validate, and illustrate an innovative concept of \textit{data-augmented} IR enhancing a numerical-IR-based state predictor with an NN. Instead of designing another numerical IR, state predictor, or its NN-based counterpart, we exploit the developed and thoroughly analyzed IRs and supplement them with a pre-trained NN classifier that is able to select the ``best'' IR for the considered task and working conditions at each time instant. The data-augmented IR, thus, combines the desired properties of the \textit{mathematics-informed} IR, such as error analysis and convergence rate and \textit{data-informed} IR with an ability to find hidden relations between working conditions and IR performance.

In this paper, we design the data-augmented IR based on the \textit{selection} of the most suitable IR by the NN for the CKE solution in actual working conditions.
The stress is laid on the descriptor specification and analysis of the proposed data-augmented integration rule error. The proposed IR \textit{minimizes} the user or designer interaction to obtain sufficiently precise IR error estimates, which, in the end, improves the accuracy of the state prediction with minimal impact on computational complexity. 

The proposed concept is analyzed and illustrated using the popular \textit{grid-based deterministic IRs} and the state prediction using the \textit{point-mass method} (PMM) widely used in signal processing and navigation applications \citep{Be:99,SiKraSo:06,LiHeGuKa:16,JePaPa:18,DuSoVeStHa:19}. It can be, however, extended for any other numerical methods. Source code in \MATLAB of the data-augmented IR is publicly available.


\section{System Definition and State Prediction}
A system with dynamics described by the discrete-time nonlinear stochastic difference state equation 
\begin{align}
	\bfx_{k+1}&=\bff_k(\bfx_k,\bfu_k)+\bfw_k\label{eq:x}
\end{align}
is considered, where $\bfx_k\in\real^{n_x}$ is the sought state of the system, $\bfu_k\in\real^{n_x}$ is the known input, and $\bfw_k\in\real^{n_x}$ is the unknown state noise at time $k$. The nonlinear function $\bff_k:\real^{n_x\times n_u}\rightarrow\real^{n_x}$ is known as well as the PDF\footnote{For the sake of notational convenience, a PDF $p_{\bfw_k}(\bfw_k)$ is denoted as $p(\bfw_k)$, if it does not lead to ambiguity.} of the noise  $p(\bfw_k)$.  

Given ``initial\footnote{In the state estimation concept, the initial PDF can be either the filtering PDF or predictive PDF from previous time instant.}'' PDF $p(\bfx_k)$, the ``predictive'' PDF $p(\bfx_{k+1})$ can be computed by the CKE as
\begin{align}
	p(\bfx_{k+1})&=\int p(\bfx_{k+1}|\bfx_{k})p(\bfx_{k})d\bfx_{k},\label{eq:cke}
\end{align}
where $p(\bfx_{k+1}|\bfx_{k})=p_{\bfw_k}\left(\bfx_{k+1}-\bff_k(\bfx_k,\bfu_k)\right)$ is the transition PDF obtained from \eqref{eq:x}. 

Considering the nonlinear function $\bff_k$ or non-Gaussian PDFs $p(\bfx_k)$ and $p(\bfw_k)$, the CKE \eqref{eq:cke} is not analytically tractable and an approximate solution, numerical PMM based on the midpoint IR in this case, is used.

\subsection{CKE Solution by Numerical Integration Rules}
\begin{figure}
	\centering
	\includegraphics[width=0.8\linewidth]{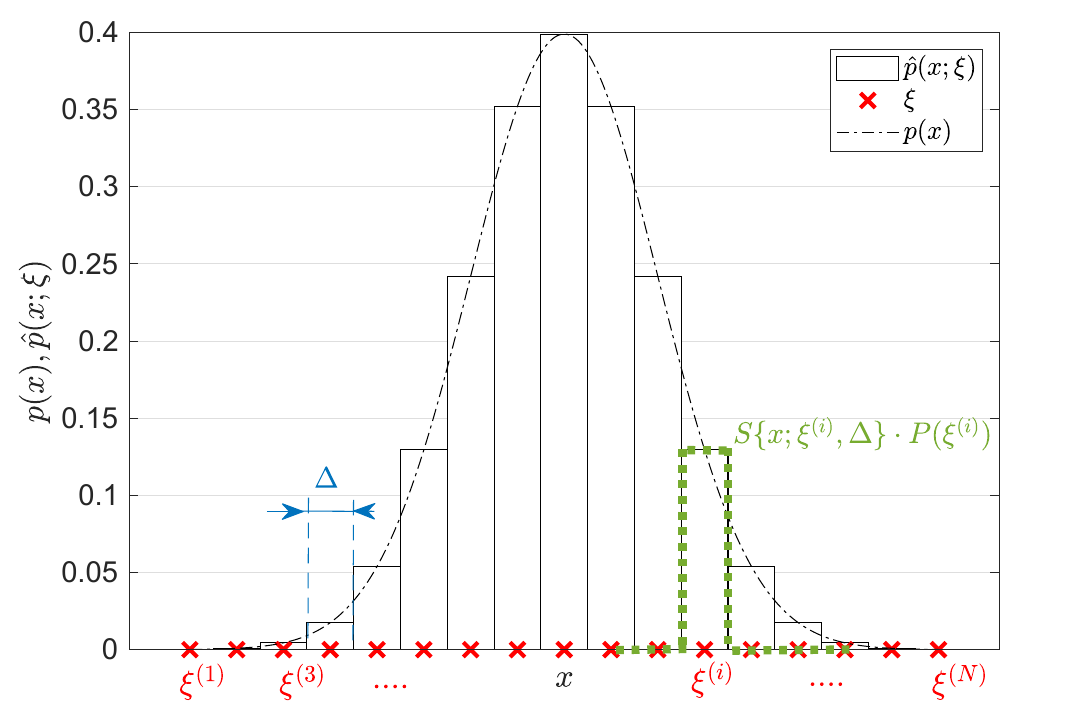}
	\caption{Illustration of point-mass PDF approximation (grid points - red, grid point neighborhood - blue, selection function - green).}
	\label{fig:pm_pdf}
\end{figure}

	 \textit{Standard} numerical solution to the CKE \eqref{eq:cke} by the PMM starts with the approximation of the initial PDF $p(\bfx_k)$  by the \textit{piece-wise} constant point-mass density (PMD) \citep{Be:99,SiKraSo:06,DuSoVeStHa:19} defined at the set of the discrete grid points $\boldsymbol{\Xi}_k=\{\boldsymbol{\xi}^{(i)}_k\}_{i=1}^N, \boldsymbol{\xi}^{(i)}_k\in\real^{n_x}$ (in a rectangular and equidistant grid), which are in the center of their respective non-overlapping neighborhood $\bfDelta_k^{(i)}$
	\begin{align}
 	p(\bfx_k)\approx\hat{p}(\bfx_k;\boldsymbol{\Xi}_k)\triangleq\sum_{i=1}^Np_{k}(\boldsymbol{\xi}^{(i)}_k)S\{\bfx_k;\boldsymbol{\xi}^{(i)}_k,\bfDelta_k\},\label{eq:pdf_pm}
	\end{align}
	where $N = N_1 \cdot N_2\ ... \cdot N_{n_x}$, and $N_i$ is a number of discretization points in $i$-th dimension of the state $\bfx_k$, $p_{k}(\boldsymbol{\xi}^{(i)}_k)$ is the value of the PDF $p(\bfx_k)$ evaluated at the $i$-th grid point $\boldsymbol{\xi}^{(i)}_k$ further also called as a \textit{weight} (the PMD is normalized to integrate to 1), $\bfDelta_k^{(i)}$ is rectangular grid cell centered at $\boldsymbol{\xi}^{(i)}_k\in\real^{d}$ with the volume $\delta_k$, where $\hat{p}(\bfx_k;\boldsymbol{\Xi}_k)$ is constant, and $S\{\bfx_k;\boldsymbol{\xi}^{(i)}_k,\bfDelta_k\}$ is an indicator function that equals to 1 if $\!\bfx_k \in \bfDelta_k^{(i)}$. Illustration of the point-mass PDF approximation \eqref{eq:pdf_pm} for \mbox{$n_x=1$} with omitted time indices is shown in Fig. \ref{fig:pm_pdf}.

Then, a significant part of the predictive PDF $p(\bfx_{k+1})$ support\footnote{The predictive PDF support can be found using a fast linearization based on the Taylor expansion or the unscented transform \citep{AnMo:79,JuUhl:04,Be:99}.} is covered by a set of the \textit{predictive} grid points $\boldsymbol{\Xi}_{k+1}=\{\boldsymbol{\xi}^{(j)}_{k+1}\}_{i=1}^N$ and the predictive PMD solving \eqref{eq:cke} becomes
\begin{align}
	\hat{p}&(\bfx_{k+1};\boldsymbol{\Xi}_{k+1})=\sum_{j=1}^N p_{k+1}(\boldsymbol{\xi}^{(j)}_{k+1})S\{\bfx_{k+1};\boldsymbol{\xi}^{(j)}_{k+1},\bfDelta_{k+1}\},\label{eq:predPMD}
\end{align} 
where
\begin{align}
	p_{k+1}(\boldsymbol{\xi}^{(j)}_{k+1})&=\int
	p(\boldsymbol{\xi}^{(j)}_{k+1}|\bfx_k)p(\bfx_{k})d\bfx_{k}.\label{eq:cke_point}
\end{align}

\subsection{Numerical Integration and its Error}
Considering the initial PDF $p(\bfx_{k})$ in the form of the PMD \eqref{eq:pdf_pm}, the integral CKE \eqref{eq:cke_point} is typically evaluated using the standard \textit{midpoint} IR given by \citep{Be:99}
\begin{align}
	\hat{P}^\mathrm{M}_{k+1}(\boldsymbol{\xi}^{(j)}_{k+1})=\sum_{i=1}^N p(\boldsymbol{\xi}^{(j)}_{k+1}|\bfx_k=\boldsymbol{\xi}^{(i)}_{k})P_{k|k}(\boldsymbol{\xi}^{(i)}_k)\delta_k.\label{eq:predProb}
\end{align}

The standard approach uses the \textit{Richardson} extrapolation of the midpoint rule \citep{HeSt:17} to achieve higher integration accuracy. The extrapolation calculates the midpoint IR twice while changing integration node density, i.e., $\boldsymbol{\Delta}_k$. Let $K(\boldsymbol{\Xi}_k;\bfDelta_k)$ be an integration rule with integration nodes distance $\bfDelta_k$ (e.g., \eqref{eq:predProb}), and $K(\boldsymbol{\Xi}_k;2\bfDelta_k)$ with doubled distance, i.e., every second grid point is used for the calculation. Then, the value of the predictive PDF at $j$-th grid point reads
\begin{align}
	\hat{P}^\mathrm{R}_{k+1}(\!\boldsymbol{\xi}^{(j)}_{k+1}\!)\!=\!   K(\boldsymbol{\Xi}_k ; \bfDelta_k\!) \!+\! \tfrac{1}{2^o-1}\left[ K(\boldsymbol{\Xi}_k ; \bfDelta_k\!) \!-\! K(\boldsymbol{\Xi}_k;2\bfDelta_k) \right]\! ,\label{eq:predProbRich1}
\end{align}
where $o=2$ is the order of the midpoint IR. 

Except for a few cases, the output of all numerical IRs, e.g., of \eqref{eq:predProb}, \eqref{eq:predProbRich1}, is subject to an error. Thus, the value of the predictive PDF \eqref{eq:cke_point} can be rewritten as
\begin{align}
	p_{k+1}(\boldsymbol{\xi}^{(j)}_{k+1}) = \hat{P}(\boldsymbol{\xi}^{(j)}_{k+1}) + \varepsilon^\mathrm{}_{k+1}(\boldsymbol{\xi}^{(j)}_{k+1}),\label{eq:IRerror}
\end{align}
where $\varepsilon^\mathrm{}_{k+1}(\boldsymbol{\xi}^{(j)}_{k+1})$ is an IR-induced error and $\hat{P}(\boldsymbol{\xi}^{(j)}_{k+1})$ is an approximation of integral \eqref{eq:cke_point} computed by a chosen IR, e.g., \eqref{eq:predProb} , \eqref{eq:predProbRich1}.

\subsection{Motivation Example and Goal of the Paper}
The midpoint IR with the Richardson interpolation generally overcomes the standard midpoint rule (regarding the maximal error). However, for certain cases, the midpoint rule can have an actual error smaller than the interpolated IR, making the rule selection inherently complicated. This situation is illustrated using a system with \textit{non-Gaussian} initial PDF and linear dynamics \eqref{eq:x} 
\begin{align}
	f_k(x_k)=Fx_k\label{eq:initPDF_motivation0}
\end{align}
where $F=1$, $n_x=1$, and $p(w_k)=\calN\{w_k;0,Q\}$, with $Q=2$. The notation $\calN\{\bfx;\hat{\bfx},\bfSigma_x\}$ stands for the Gaussian PDF of a random variable $\bfx$ with the mean $\hat{\bfx}=\mean[\bfx]$ and covariance matrix $\bfSigma_x=\cov[\bfx]$. Further, let the initial PDF be in the form of the Gaussian sum (GS) PDF
\begin{align}
	p(x_k)=\sum_{i=1}^G\alpha_k^{(i)}\calN\{x_k;\hat{x}_k^{(i)},\Sigma_{x,k}^{(i)}\},\label{eq:initPDF_motivation}
\end{align}
where $\alpha_k^{(i)}$ is a weight of the $i$-th element $\calN  \{x_k;\hat{x}_k^{(i)},\Sigma_{x,k}^{(i)}\}$ conveniently abbreviated as $\calN_k^{(i)}\{x_k\}$. The initial and predictive grids $\boldsymbol{\Xi}_k$ and $\boldsymbol{\Xi}_{k+1}$, respectively, have been designed with $N=30$ grid points covering part of the state-spaces $\calS_k=[\hat{x}_k-\sigma S_{x,k},\hat{x}_k+\sigma S_{x,k}]$ and $\calS_{k+1}=[\hat{x}_{k+1}-\sigma S_{x,k+1},\hat{x}_{k+1}+\sigma S_{x,k+1}]$, respectively, where $S_{x,k}=\sqrt{\Sigma_{x,k}}$ and $\sigma=6$. 

In total $M=6\times10^4$ Monte-Carlo (MC) simulations have been performed and the value of the predictive PDF $p(\bfx_{k+1})$ at the grid middle point $\boldsymbol{\xi}^{(j)}_{k+1}$ with\footnote{The considered grid point with $j=15$ lies at the center of the grid $\boldsymbol{\Xi}_{k+1}$ and is often associated with a significant value of the PDF. Thus, a non-negligible difference between the outputs of different IRs can be observed, which is convenient for the motivation. Selection of the particular point does \textit{not} affect the approach's generality, as illustrated later in numerical experiments.} $j=15$ has been computed
\begin{itemize}
	\item Exactly, denoted as $p(\boldsymbol{\xi}^{(15)}_{k+1})$,
	\item Numerically using the midpoint rule \eqref{eq:predProb},
	\item Numerically using the midpoint rule with the Richardson extrapolation \eqref{eq:predProbRich1} with $o=2$.
\end{itemize}
The properties of the initial PDF \eqref{eq:initPDF_motivation} were generated randomly using MC simulations, namely
\begin{itemize}
	\item Number of terms $G$ from the discrete uniform distribution with lower bound one and upper bound ten, i.e., from $\calU\{1,10\}$,
	\item Gaussian term mean $\hat{\bfx}^{(i)}$ from continuous uniform density $\calU\{-5,5\}$,
	\item Gaussian term variance $\bfP_x^{(i)}$ from continuous uniform density $\calU\{0.1,1\}$.
\end{itemize}
The performance of the numerical solutions has been validated using three criteria, namely
\begin{itemize}
	\item Root-mean-square-error
	\begin{align}
		\mathrm{RMSE}=\sqrt{\tfrac{1}{M}\sum_{m=1}^{M}\left({\varepsilon}^\mathrm{}_{k+1}(\boldsymbol{\xi}^{(15)}_{k+1})\right)^2},  \label{eq:rmse}
	\end{align}
	\item Mean absolute relative error
	\begin{align}
		\mathrm{MARE}=\tfrac{1}{M}\sum_{m=1}^{M}\abs\left(\tfrac{{\varepsilon}^\mathrm{}_{k+1}(\boldsymbol{\xi}^{(15)}_{k+1})}{p(\boldsymbol{\xi}^{(15)}_{k+1})}\right),\label{eq:mare}
	\end{align}
	where the function $\abs(\cdot)$ means the absolute value,
	\item Superiority of the considered IRs, meaning the portion of the MC simulations in which given IR provides smaller RMSE.
\end{itemize}

\begin{table}
	\caption{Accuracy of baseline IRs.}\label{tab:motivation}\vspace*{0cm}
	\begin{center}
		\setlength\extrarowheight{1pt}    
		\begin{tabular}{lccc}\hline
			& Midpoint & Richardson & Best sel.     \tabularnewline\hline
			RMSE $\times10^{-3}$ & $6.61$ & $13.7$ & $4.85$   \tabularnewline
			MARE [$\%$] & 6.1 & 19.3 & 4.9 \tabularnewline 
			Superiority & 46082 & 13918 & -- \tabularnewline\hline 
		\end{tabular}\vspace*{-0mm}
	\end{center}
\end{table}

The midpoint rule and Richardson interpolation results can be found in Table~\ref{tab:motivation}. Besides the results for the midpoint and extrapolated midpoint IRs, the ``\textit{best}'' \textit{selective} IR is given, which indicates the accuracy when the best rule is selected in each MC simulation. This best selective IR is a theoretical one, which can be computed if the exact value $p(\boldsymbol{\xi}^{(15)}_{k+1})$ \eqref{eq:cke_point} is known. The results indicate that the IR choice significantly affects the integration accuracy, and the integration error can be reduced with a suitable selection of the IR.  It is worth noting that the theoretically better (Richardson) interpolation IR provides an error smaller than the simple midpoint rules in ca. a quarter of MC simulations only. The MC differs in the shape of the initial PDF only, and there is no theoretical justification for this behavior (or prediction of it), which would allow the selection of the best rule. 

\vspace*{1em}
It means that the IR selection is \textit{not} a trivial task (it depends on multiple characteristics of the input variables), and it is problem-specific. Selection of the IR based on the IR maximum error can also be misleading as it is typically based on assessment of the maximal error, whereas the actual error might be very different.  Thus, it appears that the selection of appropriate IRs can be addressed by machine learning techniques, namely the NNs, which might extract some hidden dependencies between the model \eqref{eq:initPDF_motivation0}, initial PDF \eqref{eq:initPDF_motivation}, and the IR rule error.  

The \textit{goal} of the paper is to propose and validate a novel concept of the data-augmented IR design for enhancement of the PMM prediction by the NN-based IR selection. 

\section{Data-augmented IR Selection}

\begin{figure}
	\centering 
	\includegraphics[width=0.7\linewidth]{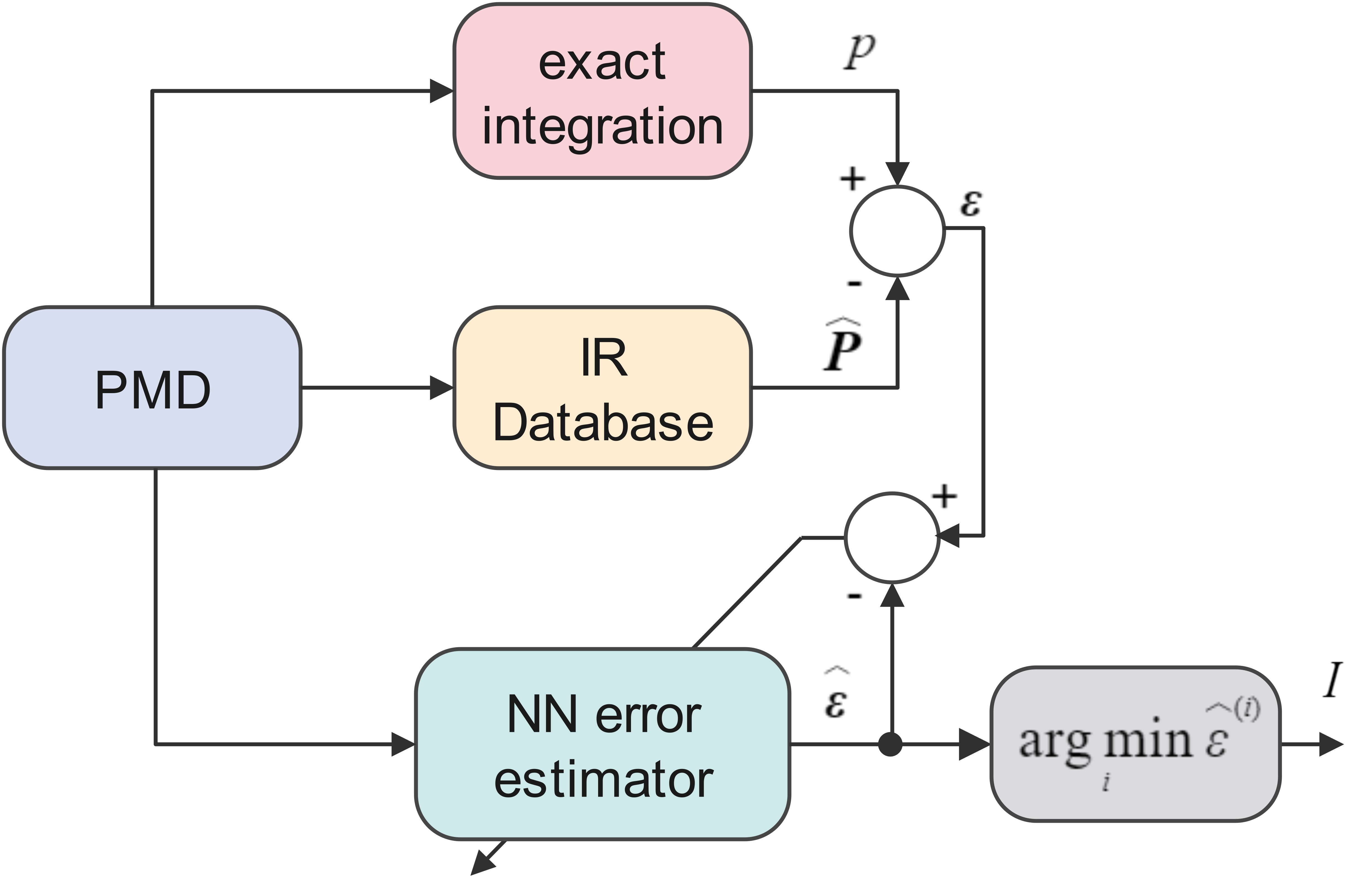}
	\caption{Training process of NN error estimator.}
	\label{fig:NNerror}
\end{figure}

The main motivation behind this approach is to consider the error magnitude 
of the IRs when designing the classifier. The key point is first to find an estimate of the IR error using NNs, as shown in Fig.~\ref{fig:NNerror}. Subsequently, an optimal IR is selected from $B$ candidate IRs (e.g., midpoint and medpoint with the Richardson extrapolation) at each time instant according to
\begin{align}\label{eq:class_optim2}
	\calB^{*} = \operatorname*{arg\,min}_b \abs\left({\hat{\varepsilon}^{(b)}_{k+1}(\boldsymbol{\xi}_{k+1}^{(j)})}\right), b=1,2,\ldots,B,
\end{align}
where instead of the true IR error ${\boldsymbol{\varepsilon}}^{(b)}_{k+1}(\boldsymbol{\xi}_{k+1}^{(j)})$ its NN-based estimate $\hat{\boldsymbol{\varepsilon}}^{(b)}_{k+1}(\boldsymbol{\xi}_{k+1}^{(j)})$ is used. 
The value of the integral calculated using the rule selected by the NN is denoted as $P_{k+1}^{\mathrm{S}}(\boldsymbol{\xi}^{(j)}_{k+1})$.
The error \textit{estimator} can be trained using supervised learning to \textit{minimize} the mean square error of the difference between the IR error ${\varepsilon}^\mathrm{}_{k+1}(\boldsymbol{\xi}^{(j)}_{k+1})$ and its estimate $\hat{\varepsilon}^\mathrm{}_{k+1}(\boldsymbol{\xi}^{(j)}_{k+1})$.

\subsection{NN Structure and Training Features}
As indicated by the analysis, the classification task is a non-trivial problem. Since no single form of the classifier is suitable for all data sets, an extensive set of methods has been developed during the last decades, including naive Bayesian, support vector machines, decision trees, and NNs \citep{Sa:18}. A NN-based classifier will be preferred due to its high versatility, efficiency, and performance.

An NN is defined by its structure, i.e., number of neurons, activation function type, number of layers, and topology \citep{Ha:99}. Unfortunately, there is no fail-proof procedure for choosing the structure of the NN, and this option depends on the problem at hand. 
Thus, an appropriate choice of the NN structure will be described for a specific example in Section~\ref{sec:num_illus}.

Independently of the NN structure, training features and evaluation criteria must be carefully selected. Principally, two different sets of features describing the initial PMD $\hat{p}(\bfx_k;\boldsymbol{\Xi}_k)$ \eqref{eq:pdf_pm} were considered:
\begin{itemize}
	\item \textit{Statistical}, characterising the PMD by the number of grid points $N$ and a set of raw and central moments,
	\item \textit{Analytical}, characterizing the PMD by the \textit{first} and \textit{second}-order central differences at considered grid points.
\end{itemize}
Both sets were supplemented with the PMD $\hat{p}(\bfx_k;\boldsymbol{\Xi}_k)$, grid points $\boldsymbol{\Xi}_k$, and the predictive grid point $\boldsymbol{\xi}^{(j)}_{k+1}$. Numerical experiments revealed that the statistical characteristics of the PMD have little effect on NN training and decision; thus, the analytical ones are preferred in this paper. The analytical features describe the \textit{spatial} behavior of the PMD, which significantly affects the IR accuracy.

Regarding the evaluation criteria, the \textit{classifier} was trained using cross-entropy as the loss function for the classification task with $B$ mutually exclusive classes. The NN classifier output is the index $\calB$ of the selected IR.


\subsection{Data-augmented IR Error Properties and NN Training}
The proposed data-augmented IRs generally benefit from the well-developed theory of stability, convergence, and error properties of the mathematics-driven IRs \citep{RaRa:78,Zwi:03}. Such property might be essential, e.g., from a certification perspective of a safety-critical system. Moreover, the data-augmented approach can also be used if no or a limited number of data is available. In such a case, the approach simply becomes the mathematics-driven one. The worst-case properties are driven by the least accurate numerical IR properties.  The NN classifier needs to be repeated if a new IR is considered. 

\section{Numerical Illustration} \label{sec:num_illus}
The performance of the NN-enhanced PMM prediction is illustrated using the set-up defined in the motivation by \eqref{eq:initPDF_motivation0}, \eqref{eq:initPDF_motivation}. The linear model with the GS initial PDF allows \textit{exact} integral evaluation and, thus, exact calculation of the predictive PDF. 

\subsection{IR Selection}
The experiment starts with a classification task where the best IR is selected out of the standard midpoint IR \eqref{eq:predProb} and the midpoint IR with the Richardson extrapolation \eqref{eq:predProbRich1}. 


The details of the training process can be summarized as: 

\begin{itemize}
	\item \textbf{Data collection}: The data was generated using $10^6$ evaluations of the integral ``true'' $p(\boldsymbol{\xi}^{(j)}_{k+1})$ \eqref{eq:cke_point} and two numerical approximations $P_{k+1}^{\mathrm{M}}(\boldsymbol{\xi}^{(j)}_{k+1})$ \eqref{eq:predProb} and $P_{k+1}^{\mathrm{R}}(\boldsymbol{\xi}^{(j)}_{k+1})$ \eqref{eq:predProbRich1}. 
	
	\item \textbf{Data pre-selection}: To improve the convergence rate of the NN training process and performance of the NN classifier/estimator, the data observations were pre-selected and only important ones were kept, i.e., those fulfilling
		${\varepsilon}^\mathrm{}_{k+1}(\boldsymbol{\xi}^{(j)}_{k+1})>\beta,$
	where $\beta$ is a threshold derived from the expected or typical magnitude of the processed initial PDF $p(x_k)$. In this experiment, $\beta=10^{-2}$ is approximately $5\%$ of the maximum value of the observed PDFs. Compensation for lower IR errors does not bring any benefit. Then, the IR value can be classified/compensated if the IR error is greater than the threshold. 
	\item \textbf{Data cross-validation}: Data was divided into a training set (80\%) and a test set (20\%) for validation.
	\item \textbf{NN structure selection}: 
 A feed-forward network with 87 identical inputs as in the previous case, followed by three fully connected layers with 128, 64, and 2 neurons with hyperbolic tangent activation functions, where the output of the last layer provides estimates of both IR errors. In this case, the total number of parameters is 19650.
 

	
	
	\item \textbf{NN parameter optimization}: The NN parameters were optimized using stochastic gradient descent with momentum (SGDM) algorithm with default/recommended learning rate parameters. 
\end{itemize}

The pre-trained NN classifiers were used for the integral evaluation, where additional $10^4$ observations were generated. The results can be seen in Table \ref{tab:illustration} w.r.t. RMSE \eqref{eq:rmse}, MARE \eqref{eq:mare}, and accuracy, which gives a ratio of correct determination of the IR with a lower error.

\begin{table}
	\caption{Selective IR Accuracy (Tab. \ref{tab:motivation} cont'd).}\label{tab:illustration}\vspace*{0cm}
	\begin{center}
		\setlength\extrarowheight{1pt}    
		\begin{tabular}{lcc}\hline
			& Best & Selective IR      \tabularnewline\hline
			RMSE $\times 10^{-3}$ & $4.85$ & $4.98$    \tabularnewline
			MARE [$\%$] & 4.9 & 5.0  \tabularnewline 
			Accuracy [$\%$]  & 100  & 95.8 \tabularnewline\hline 
		\end{tabular}\vspace*{2mm}
	\end{center}
\end{table}

The table indicates that the selective IR is only slightly worse than the best selective IR, which takes advantage of the availability of the true value of the integral. 
Analysis of this behavior revealed that the Richardson extrapolation leads to large errors in certain cases. Thus, a correct classification is required in such cases (i.e., not selection of the Richardson extrapolation). Regarding the computational complexity, Based on the simulation results, evaluation of the trained NN for the IR selection, including the features calculations, is roughly the same as evaluation of the mathematics-driven IR.

\subsection{PMM Prediction Step Error}
In the previous parts, the predictive PDF was evaluated at a single point $\boldsymbol{\xi}^{(j)}_{k+1},j=15$, only. However, to compute the predictive PMD, the convolution  \eqref{eq:predPMD} has to be evaluated $\forall j$. For this purpose, the single NN for the midpoint rule has been retrained for all possible grid points $\boldsymbol{\xi}^{(j)}_{k+1}$ with the same settings as specified in the motivation example in \eqref{eq:initPDF_motivation0}, \eqref{eq:initPDF_motivation}. The standard and NN-based compensated midpoint IRs were validated using MARE criterion \eqref{eq:mare}. The midpoint rule led to a criterion value of 16.2\%, whereas \textbf{selective IR led to} value of 6.1\%.

\subsection{Robustness, Error Compensation, and Implementation}
Further results and analyses on NN training and classification, estimator prediction step performance, and selective IR robustness can be found in \citep{DuKrMaBrSt:23} as well as discussion on a design of a method for error compensation of the numerical IRs based on the NN output. 

An exemplary source code in \MATLAB{} illustrating NN structure and training for IR error estimation is available. The code requires Deep Learning Toolbox\textcopyright. The source code is written to be easily modifiable in terms of considered dynamics\footnote{According to the simulations, the data-augmented IR improvement is not affected by the state dynamics.} (linear and nonlinear) and the state noise properties (\url{https://idm.kky.zcu.cz/files}).



\section{Concluding Remarks}
The paper dealt with a numerical solution to the Chapman-Kolmogorov equation by the point-mass method. The stress was laid on improving the numerical integration schemes by properly trained neural networks. The concept of the data-augmented, i.e., data and mathematics theory based,  IR was proposed, where the best IR for the actual set-up is selected based on the trained NN. The concept offers a significant improvement of the integration (and thus estimation) performance with rather a negligible impact on the computational complexity and allow accuracy assessment of the IR error estimate. Although the integration error estimation was discussed in the state estimation and prediction, it can be directly applied to any problem where the integral relations are solved numerically.

\textbf{Acknowledgment}: \textit{This work was co-funded by the European Union under the project ROBOPROX - Robotics and advanced industrial production  (reg. no. CZ.02.01.01\slash00\slash22\_008\slash0004590).}



\end{document}